\documentstyle[preprint,prb,aps]{revtex}
\begin{document}

\draft

\title{Unusual metallic phase in a chain of strongly interacting
particles}
\author{
E. V. Tsiper,\footnote{current address: NEC Research Institute, 4
    Independence Way, Princeton, NJ 08540}
A. L. Efros}
\address{Department of Physics, University of Utah, Salt Lake City,
Utah 84112, U. S. A.}
\date{\today}
\maketitle

\begin{abstract}
We consider a one-dimensional lattice model with the nearest-neighbor
interaction $V_1$ and the next-nearest neighbor interaction $V_2$ with
filling factor 1/2 at zero temperature.  The particles are assumed to
be spinless fermions or hard-core bosons.  Using very simple
assumptions we are able to predict the basic structure of the
insulator-metal phase diagram for this model.  Computations of the
flux sensitivity support the main features of the proposed diagram and
show that the system maintains metallic properties at arbitrarily
large values of $V_1$ and $V_2$ along the line $V_1-2V_2=\gamma J$,
where $J$ is the hopping amplitude, and $\gamma\approx1.2$.  We think
that close to this line the system is a ``weak'' metal in a sense that
the flux sensitivity decreases with the size of the system not
exponentially but as $1/L^\alpha$ with $\alpha>1$.
\end{abstract}

\newpage

The interest to the theory of one-dimensional systems is only
partially related to the study of organic conductors and other
quasi-1D compounds.  Another source of interest in the 1D physics
comes from the variety of problems which are either exactly
soluble\cite{mattis} or more amenable to computational approach.
Their solutions give guidance to intuition which can be applied to
problems in higher dimensions.

We consider a 1D system on a lattice with the following Hamiltonian:
\begin{equation}
H=J\sum_j (a^\dagger_ja_{j+1} + h.c.)+\sum_{i\neq j}V_{|i-j|}n_in_j
\label{ham}
\end{equation}
We study only the filling factor $\nu=1/2$.  In the case of the
Coulomb potential $V_{|i-j|}=1/|i-j|$ one should maintain neutrality
and change $n_i\rightarrow n_i-\nu$.

We consider the spinless fermion system at $T=0$.  One can show that
for an odd number of electrons $N$ the Hamiltonian coincides with that
for hard-core bosons.  For even $N$ the fermion-boson transformation
requires the change of periodic boundary conditions into antiperiodic.
The particle-hole symmetry can be shown to require that for even $N$
at $\nu=1/2$ the states with total quasimomenta $P$, $\pi-P$, $-P$,
and $P-\pi$ are degenerate.

The system under study undergoes structural and insulator-metal (IM)
phase transitions when the hopping amplitude $J$ is varied.  The
general point of view is that at small $J$ the ground state has a
crystalline order and is insulating.  In the free-fermion limit of
large $J$ the system does not have long-range order and is metallic.

In the case of nearest-neighbor interaction and only then the problem
is exactly soluble.\cite{yang,des,suth} In this case the structural
transition occurs simultaneously with the IM transition.\cite{suth} In
principle, two separate transitions are not forbidden.  Nevertheless,
in the qualitative arguments below we assume that these transitions
are connected to each other and occur at the same $J_c$.

We concentrate here on the IM transition in a model with the
nearest-neighbor and the next-nearest neighbor interactions, the
so-called [$V_1,V_2$]-model.  It has been studied\cite{emery} in
connection with the spin version of the Hamiltonian Eq.~(\ref{ham}).
The IM phase diagram for this model has been studied recently in
Ref.~\onlinecite{dag}.

We detect the IM transition by analyzing flux
sensitivity\cite{kohn,scal} $\delta E=|E_p-E_a|$, where $E_p$ and
$E_a$ are the ground-state energies for periodic and antiperiodic
boundary conditions.  For simplicity, we take $E_a$ to be the
lowest-energy state with the same quasimomentum $P$ as $E_p$.

Starting from the ordered phase at $J=0$ and using perturbation theory
with respect to $J$, one can show that $\delta E\sim J^N$ at small $J$
and hence falls off exponentially with the system size $L=2N$.  For
free fermions $\delta E=\pi J/L$.  Thus, the dependence of the product
$L\delta E$ on $L$ and $J$ is a nice criterion for detection of the IM
transition.  We obtain this dependence by exact diagonalization
technique.

The idea we want to check here is that the IM transition is closely
related to the point defect with the lowest energy in the crystalline
phase.  At finite $J$ the point defect forms a band.  The transition
occurs at such $J$ that the lowest edge of the band comes close to the
energy of the ground state.\cite{old} At this point the ground state
becomes a strong mixture of the crystalline and defect states.  This
mechanism reminds the idea of zero-point defectons proposed by Andreev
and Lifshitz.\cite{AL}

Such a simple picture of the transition implies that the critical
value of $J$ is determined by the energy $E_d$ of the defect at $J=0$.
The empirical rule we propose is $J_c=\beta E_d$, where $\beta$ is
some number.  For the exactly soluble problem with nearest-neighbor
interaction $\beta=0.5$.  For the Coulomb problem $E_d=2\ln2-1=0.386$.
Our computations\cite{tsiper} show that for the Coulomb interaction
$J_c$ is between 0.17 and 0.3, which gives $0.44<\beta<0.77$.  In the
2D case we have found\cite{tsiper1} that $\beta$ is approximately in
the same interval.

Using the empirical relation 

\begin{equation}
J_c=0.5E_d
\label{cond}
\end{equation}
we can construct the IM phase diagram for the [$V_1,V_2$]-model (see
Fig.~1).  Note that the explicit value of $\beta$ is not important for
the qualitative results.  We choose $\beta=0.5$ to get the correct
value of $J_c$ for the case $V_2=0$, where it is known exactly.  We
show below that this is a right choice in a wide range of $V_1$ and
$V_2$.

Two competing crystalline structures exist in the [$V_1,V_2$]-model at
$J=0$.  The structure 1 is $\bullet$$\circ$$\bullet$$\circ$, where
$\bullet$ stands for an occupied and $\circ$ stands for an empty site.
The structure 2 is $\bullet$$\bullet$$\circ$$\circ$.

Dotted lines in Fig.~1 indicate three regions.  At $J=0$ the structure
1 has the lowest energy in the region I, where
$\Delta\equiv2V_2-V_1<0$.  The lowest-energy defect in this structure
has energy $-\Delta$ and represents a shift of an electron to the
nearest site.  The structure 2 is stable in the regions II and III,
where $\Delta>0$.  In the region II the lowest defect has energy
$\Delta$ and is also a shift of one electron.  In the region III
another defect wins, which has energy $V_2$.  This defect is a
``domain boundary'', when a portion of a crystal is shifted one site
to the right or to the left.  Such shift, in fact, produces two domain
boundaries simultaneously.

Eq.~(\ref{cond}) gives the dependence $J_c(V_1,V_2)$ that is shown in
Fig.~1 with solid lines.  These lines separate insulating and metallic
phases.  To obtain $J_c(V_1,V_2)$ one should substitute into
Eq.~(\ref{cond}) the proper expression for the minimum defect energy
$E_d(V_1,V_2)$ at $J=0$ in each of three regions as discussed above.
The lower solid line shows the IM transition associated with the
crystalline structure 1.  The upper solid line shows the same
transition for the structure 2.  It consists of two straight lines in
two different regions, II and III, which correspond to the different
types of defects.

Fig.~2 shows the results of numerical computation of $L\delta E/J$ as
a function of $J$ at fixed $V_1$ and $V_2$ for a system of 14
electrons.  The data for smaller sizes are not shown.  However, they
have been used to find the critical value $J_c$ by extrapolation to
$1/L\rightarrow 0$.  At ($V_1,V_2$) equal to (1,0), (0,1), and (1,1)
our criterion predicts the transition at $J_c=0.5$; at (4,1) it
predicts $J_c=1$.  These values are indicated by the points $a$, $b$,
$c$, and $d$ in Fig.~1, and by arrows in Fig.~2.  The value $J_c=0.5$
is exact for the point $(1,0)$.\cite{yang,des,suth} The results of
extrapolation give predicted values for the first three points with a
15\% accuracy.\cite{comment} For the point (4,1) we have gotten
$J_c=1.2\pm0.1$.  Thus, we may conclude that the Eq.~(\ref{cond})
works very well in a wide range of $V_1$ and $V_2$.

The most important prediction of the phase diagram Fig.~1 is existence
of a metallic region between the solid lines which extends infinitely
for arbitrarily large $V_1$ and $V_2$ close to the line
$\Delta=2V_2-V_1=0$.  Consider the curves in Fig.~2 corresponding to
$(V_1,V_2)=(1,0.48)$ and $(1,0.52)$.  Now with changing $J$ we are
moving almost along the line $\Delta=0$ in Fig.~1.  In the first case
we deviate a little towards the Crystal 1, and in the second case ---
towards the Crystal 2.  Both lines intersect the IM phase lines at
large $V_1$, $V_2$, predicting $J_c=0.02$ in both cases.  One can see
in Fig.~2 that this prediction is basically fulfilled in the sense
that the exponential dependence on $J$ disappears near this point.
For $J>J_c$ the system, however, does not look like an ordinary metal,
where $L\delta E$ should be size independent.  In fact, we have
observed a weak dependence of $L\delta E$ on $L$ in a wide range of
$J$ between $J=J_c$ and $J\approx0.4$.

Fig.~2 also shows $\delta E$ for $(V_1,V_2)=(1,0.50)$.  Now with
decreasing $J$ we are moving exactly along the line $\Delta=0$.  In
this case the exponential transition to the dielectric phase is absent
for arbitrarily small $J$, in agreement with our phase diagram Fig.~1.
However, there is some size dependence of $L\delta E$ along the line
$\Delta=0$ in the region $J\ll1$.  It can be described as $\delta
E\sim 1/L^\alpha$ with $\alpha>1$.  Thus, it is not a regular 1D metal
where $\alpha=1$. An alternative interpretation of the same data would
be an exponential size dependence $\delta E\propto\exp(-L/\xi)$ with
anomalously large correlation length $\xi$.

Now we study more carefully the close vicinity of the line $\Delta=0$
far from the origin.  In the region $\Delta\ll V_1,V_2$ the spectrum
of energies at $J=0$ has two scales.  The large scale is determined by
$V_1$ and $V_2$, while the second scale is $|\Delta|$, which is the
energy necessary to produce a defect.  When $\Delta=J=0$ the ground
state is macroscopically degenerate.

To separate these two scales we consider a limit
$V_1,V_2\rightarrow\infty$, $J$ and $\Delta$ being finite.  In this
limit the size of the Hilbert space can be greatly reduced.  Only the
states which are degenerate at $\Delta=J=0$ should be taken into
account.  These states are such that neither three electrons nor three
holes occupy adjacent sites.

The reduction of the Hilbert space size is from $C_L^{L/2}$ to
approximately $f_{L-2}$, where $f_n$ denote the Fibonacci numbers,
defined by $f_n=f_{n-1}+f_{n-2}$, $f_0=f_1=1$.  At large $n$ one
has\cite{sloane} $f_n\approx((1+\sqrt{5})/2)^{n+1}/\sqrt{5}$.

With this reduction we can increase $L$ up to 40
($f_{38}=.63\times10^8$).  Fig.~3 shows $L\delta E/J$ as a function of
$\Delta/2J$ obtained for different $L$.  The maximum occurs not at
$\Delta=0$, as can be expected from naive consideration, but at
$\Delta/2J\approx-0.6$.  Accurate size extrapolation shown in Fig.~4
demonstrate that at this point $\delta EL/J$ stays finite as $L$ goes
to infinity.  Thus, the system at $\Delta\approx1.2J$ is a normal
metal.  The flux sensitivity in the limit $L\rightarrow\infty$ is less
than the value $\pi$ for free fermions and is equal $L\delta
E/J\approx2.5$.  In the phase diagram Fig.~1 the ``magic'' metallic
line $\Delta=1.2J$ is shown with dashed line.  This line appears,
obviously, as a result of quantum mixture of the two different ordered
phases.

Fig.~3 shows also the energy per particle as a function of $\Delta/2J$
obtained in the same limit.  We have not found any singularity in the
energy in the region of interest.  The gap between the ground and the
lowest excited states with the same total quasimomentum at the magic
metallic line scales to zero linearly in $1/L$, as shown in the inset
to Fig.~3.  Note that usually a crystalline phase on the lattice has a
finite gap.

The inset in Fig.~4 shows the reciprocal correlation length
$1/\xi=-d\ln(L\delta E)/dL$ as a function of $\Delta/2J$ as obtained
from the slopes of the curves in Fig.~4 at largest $L$.  Note that the
condition $\xi<L$ corresponds to $1/\xi>0.25$.  Thus, we have a real
exponential behavior for $-3<\Delta/2J<2$.  At large negative values
of $\Delta/2J$ the ground state of the system is the crystal with the
structure 2 with a small admixture of defects which are fragments of
the structure 1.  At large and positive $\Delta/2J$ one has the
opposite picture.  In the intermediate region the ground state is a
mixture of these two structures.  If we extrapolate $1/\xi$ in each of
the exponential regions, we find that it turns into zero approximately
at the boundaries of the metallic strip, shown by two parallel solid
lines in Fig.~1.  This is natural, since the naive picture which leads
to Fig.~1 does not take into account mixing of two crystalline
structures.

The small value of $\xi$ in the intermediate region suggests that the
size dependence of $L\delta E$ is not exponential near the magic line.
This would imply the existence of another phase, which may be named a
``weak metal.''  If such phase exists, there should be a phase lines
which separate the weak metal from the normal metal, where $L\delta E$
is size independent.  The inset in Fig.~1 shows schematically the
region of the normal metallic phase.  This diagram is similar to the
one obtained in Ref.~\onlinecite{dag}, except it predicts an infinite
metallic line in the plane ($V_1,V_2$).

Finally, we have shown that a simple rule Eq.~(\ref{cond}) provides a
reasonable description of the phase diagram of IM transition in the
[$V_1,V_2$]-model.  We have found an interesting metallic phase which
exists at any small values of $J$.  The ground state of this phase is
a mixture of two crystalline phases with moving boundaries.  The
nature of a small deviation of the metallic phase from the line
$\Delta=0$ is not clear.

We are grateful to John Worlock for reading the manuscript.  We
acknowledge support of UCSB, subcontract KK3017 of QUEST, and support
of the San Diego Supercomputer Center.

\begin{figure}
\caption{Phase diagram of [$V_1,V_2$]-model.  Solid lines show the
diagram as obtained from Eq.~(2).  The dotted lines separate regions
I, II, and III.  The point $a$ is known exactly; the points $b$, $c$,
and $d$ are checked by computations.  The long-dashed lines in the
main figure and in the inset show the ``magic'' metallic line.  The
short-dashed lines in the inset show schematically the region of the
normal metallic phase where $L\delta E$ independent of $L$.
\label{fig1}}
\end{figure}

\begin{figure}
\caption{Dependence of flux sensitivity in units of $J/L$ on $J$ for
different ($V_1,V_2$) for the system with 14 electrons as obtained by
exact diagonalization.  The arrows show the transition points
predicted by the phase diagram.  The dashed line shows the free
fermion result $L\delta E/J=\pi$.
\label{fig2}}
\end{figure}

\begin{figure}
\caption{Flux sensitivity $L\delta E/J$ for different $L$ and the
ground-state energy $E$ per particle for $L=40$ as functions of
$\Delta/2J$.  The energy $E$ is measured from the classical energy of
the crystalline structure 2.  The inset shows the excitation gap along
the magic metallic line vs. $1/L$.  All results are obtained by exact
diagonalization in the limit $V_1,V_2\rightarrow\infty$.
\label{fig3}}
\end{figure}

\begin{figure}
\caption{Size dependence of flux sensitivity for different values of
$\Delta/2J$ in the limit $V_1,V_2\rightarrow\infty$.  The inset shows
the slope $1/\xi$ as obtained from this size dependence at large $L$
vs. $\Delta/2J$.  The slope $1/\xi$ can be considered as the
reciprocal correlation length when $\xi<L\sim40$.
\label{fig4}}
\end{figure}


\begin{references}
\bibitem{comment} We claim such a high accuracy for the result of
extrapolation because the size dependence looks very similar to the
case ($V_1,V_2$)=(1,0), where exact $J_c$ is known.  In the Coulomb
case the extrapolation is more uncertain.
\end{references}
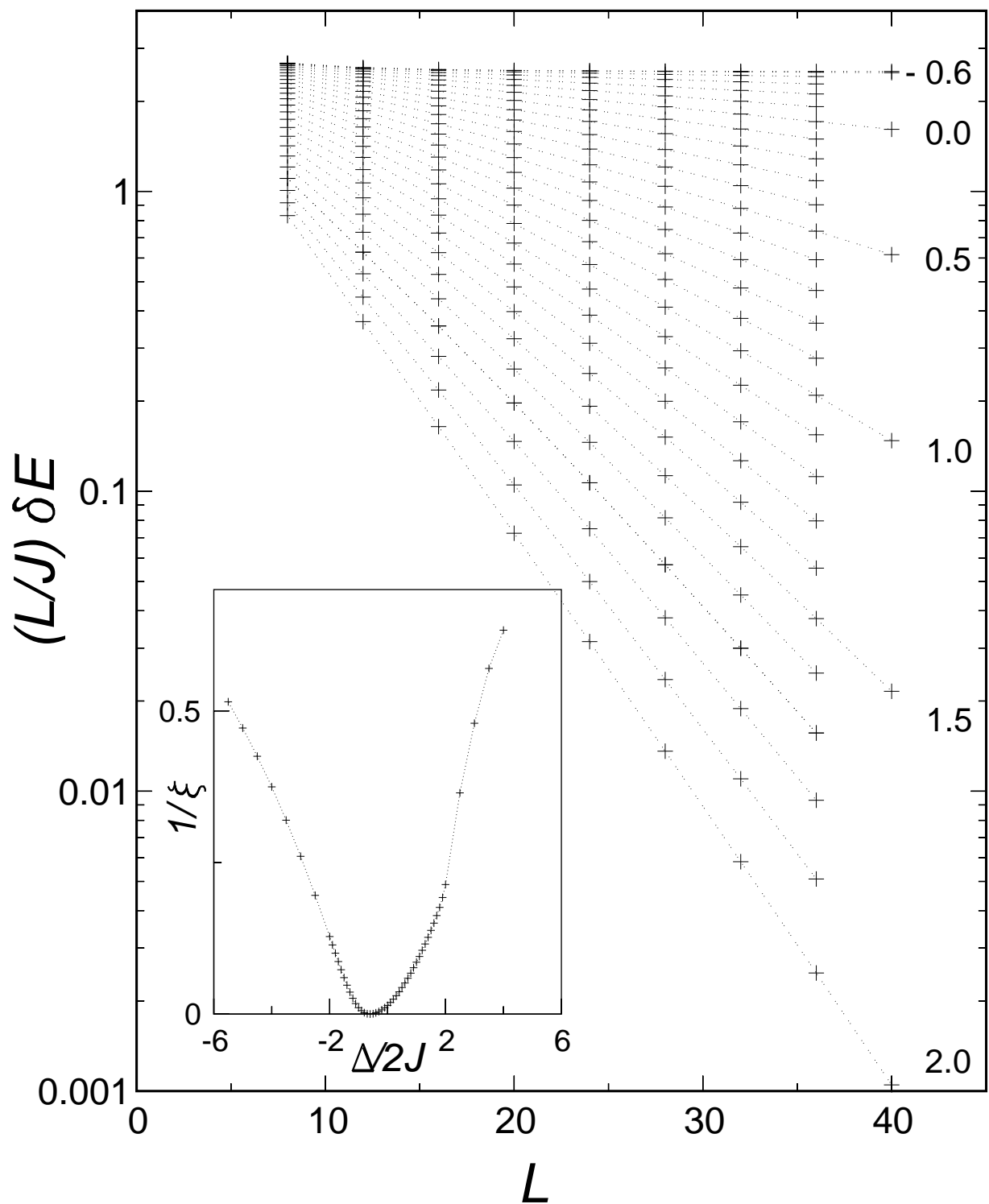
\end{document}